# Generation of shock waves from localized sources: The case of the Burgers equation

Yair Zarmi
Jacob Blaustein Institutes for Desert Research
Ben-Gurion University of the Negev
Midreshet Ben-Gurion, 8499000, Israel

Abstract
It is shown that the shock wave solutions of the Burgers equation can be generated from localized sources. The evolution equation obeyed by the sources has a novel characteristic: It has a single-soliton solution as well as an infinite family of localized-hump solutions.

E-mail: zarmi@bgu.ac.il

## 1. Introduction

The traveling wave solutions of integrable equations are often found through the use of an auxiliary function, usually denoted by $\tau$. Representations in terms of such auxiliary functions of the soliton solutions of quite a few evolution equations have been presented in the literature (see, e.g., [1-17]). However, the traditional expressions for $\tau$ do not convey any clue for the localized nature of the soliton solutions of the many evolution equations discussed in the literature. For example, the soliton solutions of the KdV equation [18] are expressed as:

$$u(t,x) = 2\partial_x^2 \log\{\tau(t,x)\} \ . \tag{1}$$

In the case of the single-soliton solution of Eq. (1), the traditional expression for $\tau(t,x)$ is

$$\tau(t,x) = 1 + e^{2k(x+4k^3 t+\delta)} \ , \tag{2}$$

where $k$ is the wave number. $\tau(t,x)$ displays a surface that varies between 1 and $\infty$, whereas the single-soliton solution is localized in the *x-t* plane.

It has been recently shown [19] that, thanks to the non-uniqueness of the function $\tau(t,x)$, the soliton solutions of quite a few evolution equations can be generated from localized sources.

A question that arises naturally is whether the same idea can be applied to traveling waves of a different structure. In this paper, the idea is extended to the case of shock waves, through the analysis of the Burgers equation,

$$u_t = 2u u_x + u_{xx} \ . \tag{3}$$

Eq. (3) was originally developed for the approximate description of the propagation of weak shock waves in one dimension in an adiabatically expanding gas [20,21]. It has been also applied as an approximate description of the dynamics of vehicle flow on highways [22-24]. Its characteristics differ from those of integrable equations, which have soliton solutions, such as the KdV, mKdV, NLS and KP II equations in two ways: It is not integrable in the Inverse-Scattering sense, and its solutions are fronts (shock waves) rather than solitons. Still, the front solutions can be constructed in terms of an auxiliary function, which is also not unique. This freedom allows for generating the front solutions from localized sources.

## 2. Localized sources for shock waves

The Hopf-Cole transformation [25-27] yields the shock wave (front) solutions of Eq. (3):

$$u(t,x) = \partial_x \log \tau(t,x) \quad . \tag{4}$$

An $N$-shock solution is generated by

$$\tau(t,x) = \sum_{i=0}^{N} e^{\theta_i} \quad , \quad \theta_i = k_i(x + k_i t + \delta_i) \quad , \quad 0 \leq k_0 < k_2 < \cdots < k_N \quad . \tag{5}$$

The boundary condition for the solution is

$$u(t,x) \underset{t,x \to -\infty}{\to} k_0 \quad . \tag{6}$$

$\tau(t,x)$ generates a surface that for $t, x \to -\infty$, starts at 0 when $k_0 > 0$ or at 1 when $k_0 = 0$, and diverges to infinity for $t, x \to +\infty$. As such, it does not provide any hint regarding the structure displayed by the solution. Figs. 1 and 2 show, respectively, $\tau(t,x)$ and $u(t,x)$ for a single-shock solution, and Figs. 3 and 4 – for a three-shock solution. In both examples, the solutions obey vanishing boundary conditions at $-\infty$.

To generate the shock-wave solutions out of localized sources, consider the function

$$\tau_E(t,x) = \tau(t,x) e^{-\frac{1}{N+1} \sum_{i=0}^{N} \theta_i} = \sum_{i=0}^{N} e^{(k_i - \langle k \rangle)x + (k_i^2 - \langle k^2 \rangle)t + k_i(\delta_i - \langle \delta \rangle)} \tag{7}$$

In Eq. (7), $<k>$, $<k^2>$ and $<\delta>$ are the mean values of the corresponding entities.

The definition of $\tau_E(t,x)$ ensures that the sum of all the exponents in Eq.(7) vanishes. Hence, in every direction in the $x$-$t$ plane, there are some positive and some negative exponents. Consequently, $\tau_E(t,x)$ grows indefinitely in all directions in the plane, so that the function,

$$S(t,x) = \frac{1}{\tau_E(t,x)} = \frac{1}{\sum_{i=0}^{N} e^{(k_i - \langle k \rangle)x + (k_i^2 - \langle k^2 \rangle) + k_i(\delta_i - \langle \delta \rangle)}} \quad , \tag{8}$$

is localized. $S(t,x)$ vanishes asymptotically for $x \to \pm\infty$ and for $t \to \pm\infty$.

In terms of $S(t,x)$, Eq. (4) is replaced by

$$u(t,x) = -\partial_x \log S(t,x) + \langle k \rangle \quad . \tag{9}$$

Thus, the $N$-shock solution is generated from a localized source.

The constant shift is required to ensure the boundary condition at $t, x \to -\infty$. It is of no fundamental importance. Transforming to a moving,

$$\begin{aligned} t' &= t \\ x' &= x + 2\langle k \rangle t \\ u(t,x) &= v(t,x') + \langle k \rangle \end{aligned} \quad , \tag{10}$$

eliminates the shift. Eq. (3) is transformed into (the prime in the definition of $t'$ and $x'$ is dropped)

$$v_t = 2v v_x + v_{xx} \quad . \tag{11}$$

$v(t,x)$ is generated by the localized source, $S(t,x)$:

$$v(t,x) = -\partial_x \log S(t,x) \tag{12}$$

While $u(t,x)$ vanishes at $t, x \to -\infty$, $v(t,x)$ tends to $k_0 - \langle k \rangle$.

For example, for a single front that obeys vanishing boundary at $-\infty$ (corresponding to $k_0 = 0$)

$$\tau_E(t,x) = 2\cosh(\theta_1/2) \quad \Rightarrow \quad S(t,x) = \frac{1}{2\cosh(\theta_1/2)} \quad , \tag{13}$$

The resulting source, $S(t,x)$, shown in Fig.5, displays a soliton along the boundary of the shock front of Fig. 2.

For $N \geq 2$, $S(t,x)$ is localized in the region, in which all fronts interact. The intuitive explanation is simple: This is the region of overlap amongst all the solitons, each of which is the source for one of the shocks. The plot of $S(t,x)$ for $N = 3$ is shown in Fig. 6.

### 3. Evolution equation of source

Exploiting the definition, Eq. (8), one finds that the source, $S(t,x)$, obeys the following evolution equation:

$$S_t = \sigma^2 S + 2\langle k \rangle S_x + S_{xx} - 2\frac{S_x^2}{S} . \tag{14}$$

Eq. (14) has a novel feature: It has a single-soliton solution (the source of the single-front solution of Eq. (3); see Eq. (13) and Fig. (5)) as well as an infinite family of localized-hump solutions (the sources of multi-front solutions of Eq. (3); see, e.g., Fig. 6).

There is a reciprocity relation between, Eq. (14) and the Burgers equation, Eq. (3). In a moving frame of reference with

$$\{t,x\} \rightarrow \{t,\xi\} \quad (\xi = x + 2\langle k \rangle t) , \tag{15}$$

and

$$S(t,x) = e^{\sigma^2 t} e^{-\int H(t,\xi) d\xi} , \tag{16}$$

$H$ obeys the Burgers equation:

$$H_t = 2 H H_\xi + H_{\xi\xi} . \tag{17}$$

### 4. Concluding comments

It has been shown that the shock waves described by the Burgers equation can be generated from localized sources. The evolution equation obeyed by the sources has a novel characteristic: It has a single-soliton solution as well as an infinite family of localized-hump solutions.

References

1. R. Hirota, Phys. Rev. Lett. **27**, 1192-1194 (1971).

2. M.J. Ablowitz, D.J. Kaup, A.C. Newell, H. Segur, Stud. Appl. Math. **53**, 249-315 (1974).

3. V.E. Zakharov, S.V. Manakov, Sov. Phys. JETP **44**, 106-112 (1976).

4. M.J. Ablowitz, H. Segur, Solitons and the Inverse Scattering Transforms, SIAM, Philadelphia, 1981.

5. S.P. Novikov, S.V. Manakov, L.P. Pitaevskii, V.E. Zakharov, Theory of Solitons, Consultant Bureau, New York, 1984.

6. A.C. Newell, Solitons in Mathematics and Physics, SIAM, Philadelphia, PA, 1985.

7. R. Hirota, Physica D **18**, 161–170 (1986).

8. R. Hirota, Y. Ohta, J. Satsuma, Progr. Theor. Phys. Suppl. **94**, 59–72 (1988).

9. R. Hirota, The Direct Method in Soliton Theory, Cambridge University Press, 2004.

10. S. Chakravarty, Y. Kodama, J. Phys. A: Math. Theor. **41**, 275209 (2008).

11. S. Chakravarty, Y. Kodama, Contemp. Math. **471**, 47-61 (2008).

12. S. Chakravarty, Y. Kodama, Stud. Appl. Math. **123**, 83–151 (2009).

13. V.E. Zakharov and A.B. Shabat, Soviet JETP **34**, 62-69 (1972).

14. J.P. Gordon, Optics Lett. **8**, 596-598 (1983).

15. S.V. Dmitriev, D.A. Semagin, A.A. Sukhorukov and T. Shigenari, Phys. Rev. E **66**, 046609 (2002).

16. J.R. Yan, L.X. Pan and J. Lu, Chinese Phys. **13**, 441-444 (2003).

17. Y. Kodama and A.V. Mikhailov, Obstacles to Asymptotic Integrability, pp. 173-204 in *Algebraic Aspects of Integrable Systems,* ed. by A.S. Fokas & I.M. Gelfand (Birkhäuser, Boston, 1997).

18. D.J. Korteweg, G. De Vries, Philos. Mag. **39**, 422-443 (1895).

19. Y. Zarmi, J. Math. Phys. **59**, 122701 (2018).

20. J.M. Burgers, *The nonlinear Diffusion equation* (Reiedl, Dordtrecht, 1974).

21. R.A. Kraenkel, J.G. Pereira, and E.C. de Rey Neto, Phys. Rev. E **58**, 2526-2530 (1998).

22. M.J. Lighthill and G.B. Whitham, Proc. Roy. Soc. London A: Math. Phys. Eng. Sci. **229**, 317-345 (1955).

23. B.S. Kerner and P. Konhäuser, Phys. Rev. E **50**, 54-83 (1994).

24. A. Takac, Proc. Appl. Math. Mech. **5**, 633-634 (2005).

25. A.R. Forsyth, *Theory of differential equations. Part IV − Partial differential equations*, (Cambridge Univ. Press, Cambridge, 1906, Republished by Dover, New York, 1959).

26. Hopf, E., Comm. Pure Appl. Math. **3**, 201-230 (1950).

27. Cole, J. D., Quart. Appl. Math. **9**, 225-236 (1951).

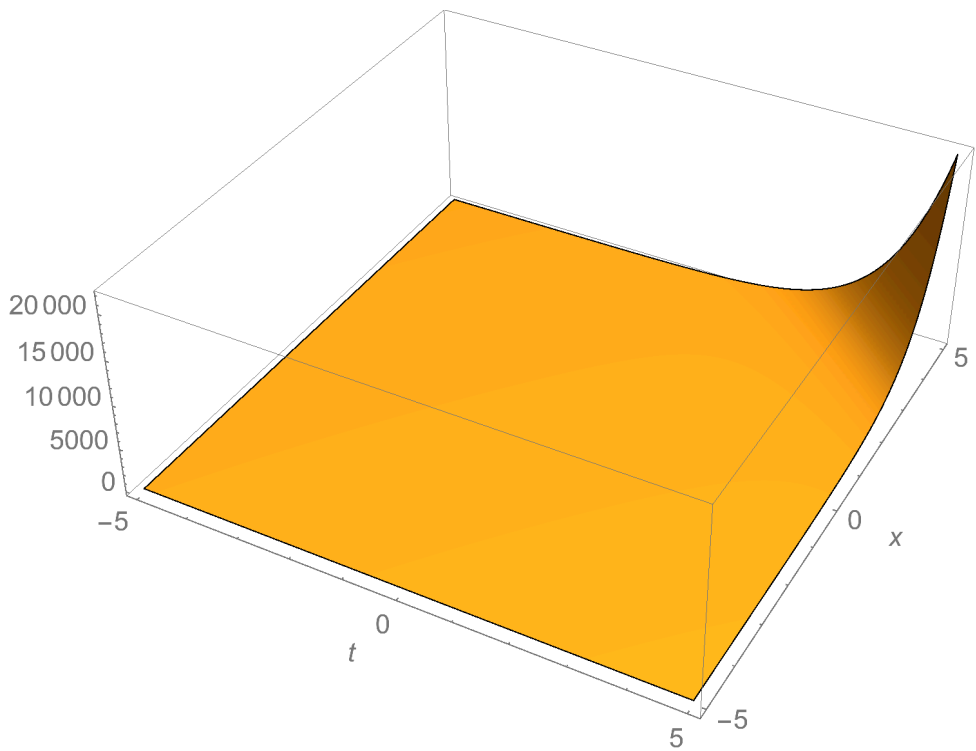

Fig. 1 $\tau(t,x)$ (Eq.(3)) for a single-shock solution ($N = 1$ in Eq. (5)). $k_0 = 0$, $k_1 = 1.$; $\delta_1 = 0$.

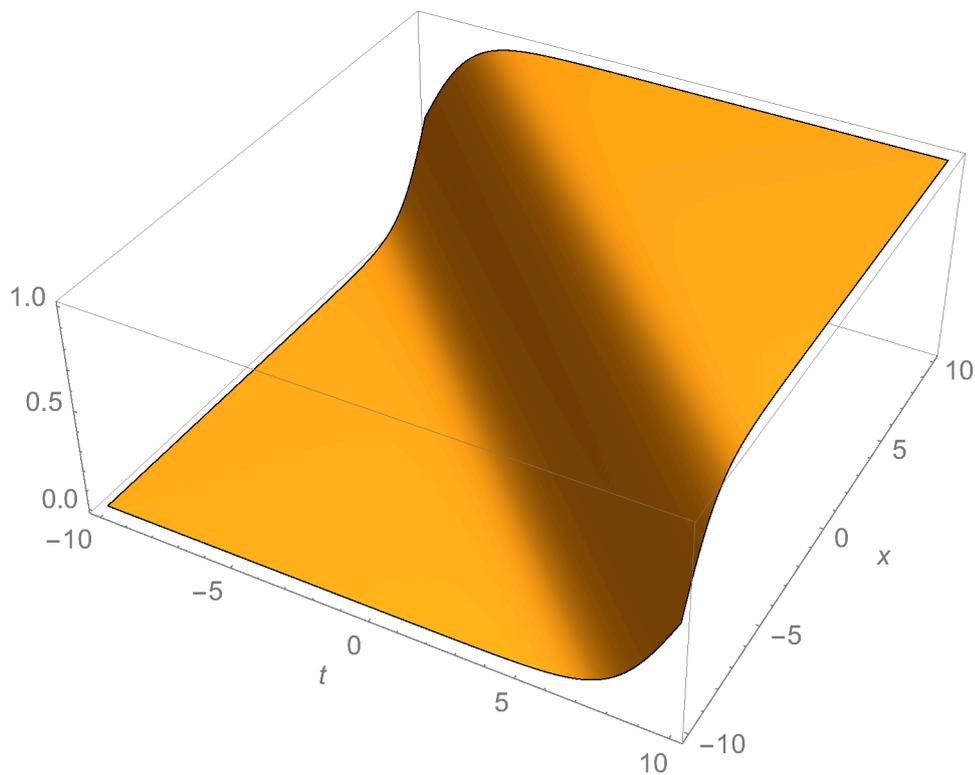

Fig. 2 Single-shock solution. Parameters as in Fig. 1.

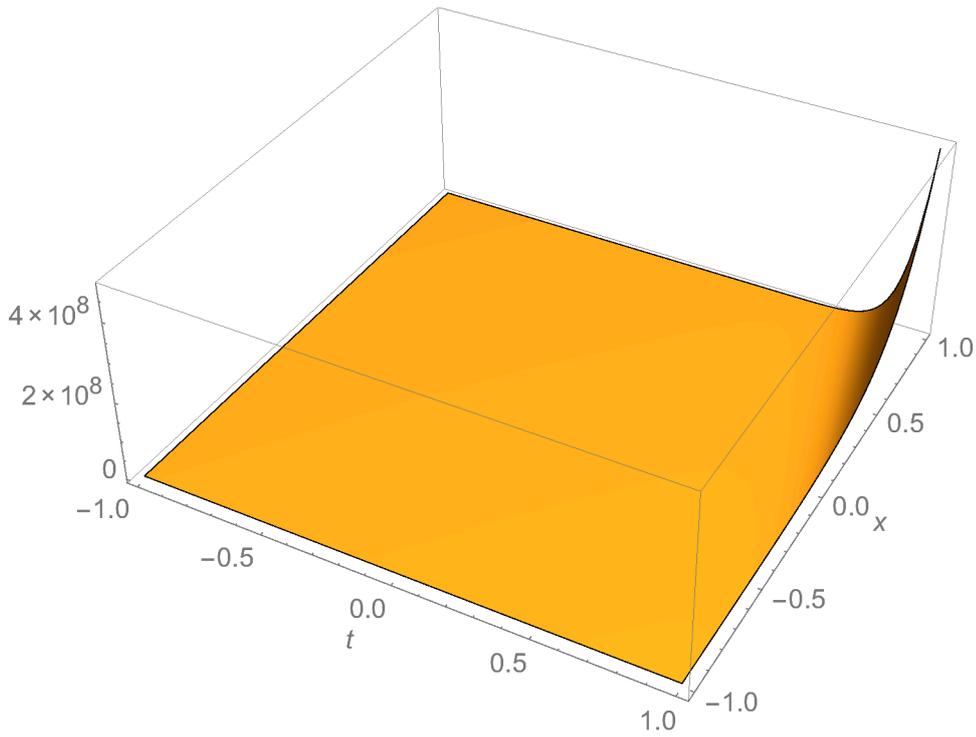

Fig. 3 $\tau(t,x)$ (Eq.(3)) for a three-shock solution ($N = 3$ in Eq. (5)).
$k_0 = 0$, $k_1 = 1.$, $k_2 = 2.$, $k_3 = 4.$; $\delta_1 = \delta_2 = \delta_3 = 0$.

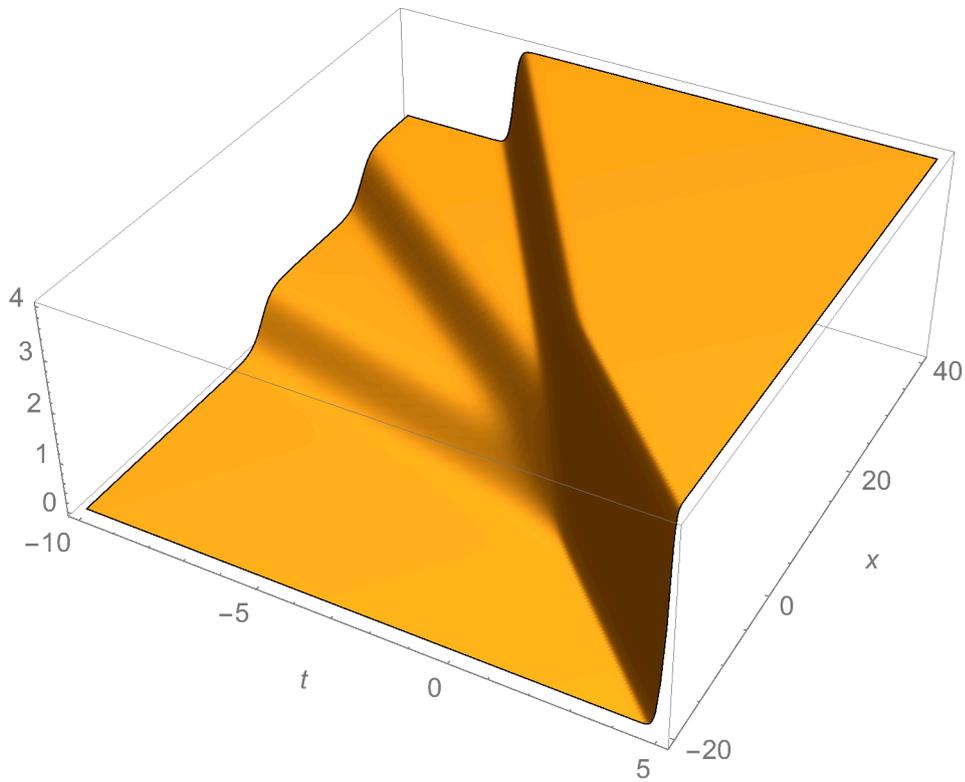

Fig. 4 Three-shock solution. Parameters as in Fig. 3.

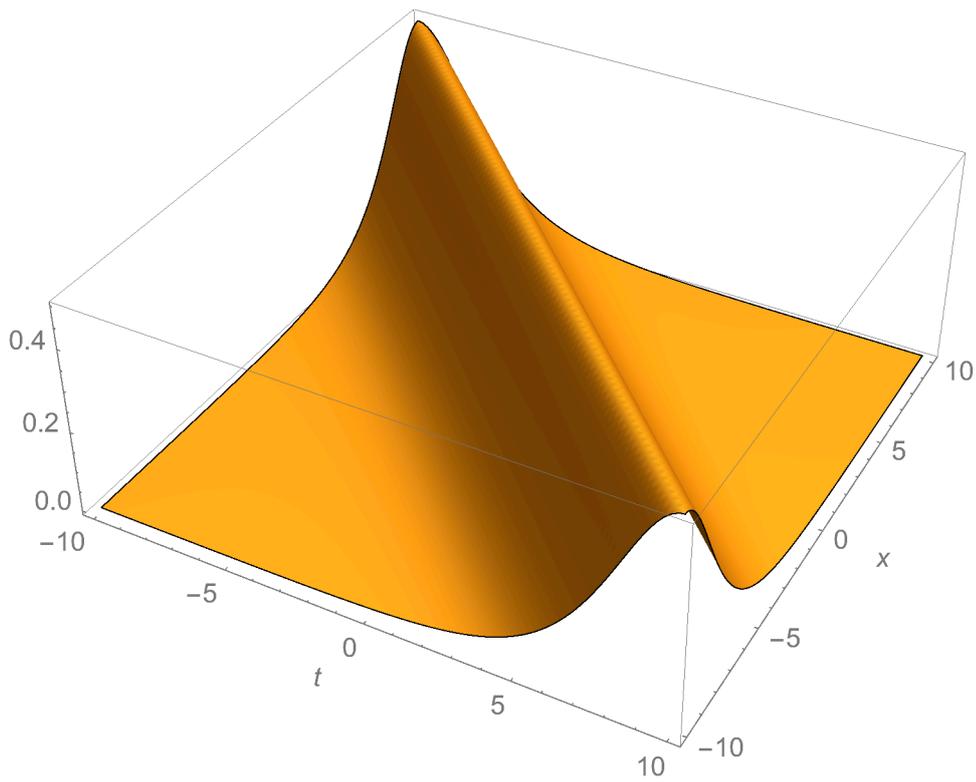

Fig. 5 Source, $S(t,x)$, for single-shock solution (Eqs. (7)-(9), (13)). Parameters as in Fig. 1.

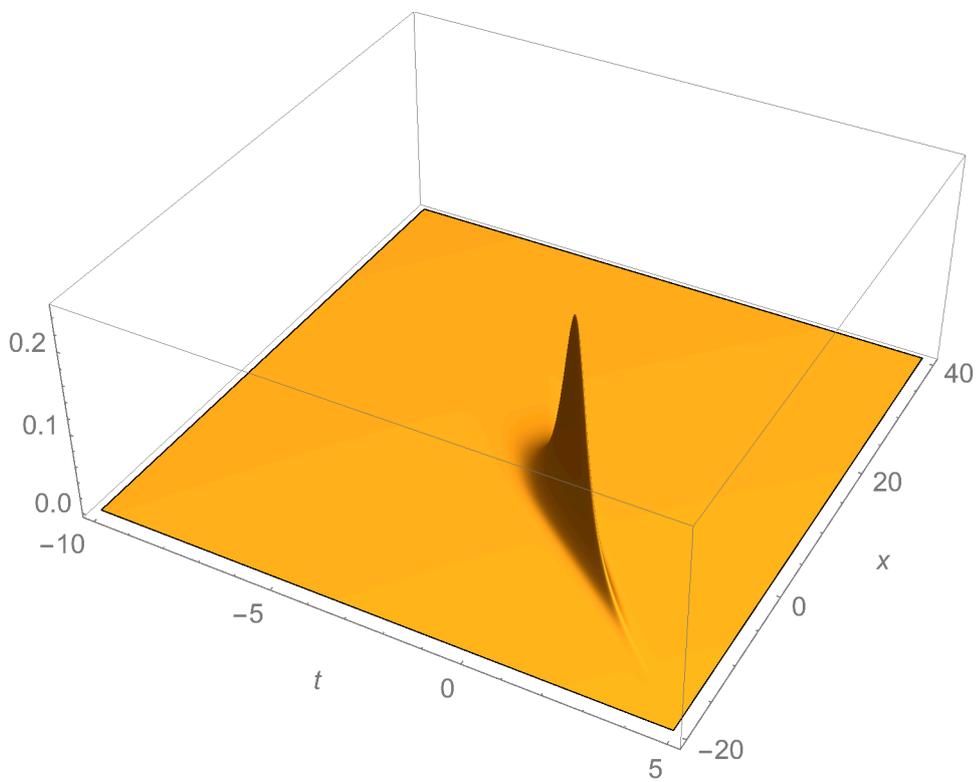

Fig. 6 Source, $S(t,x)$, for three-shock solution (Eqs. (7) - (9)). Parameters as in Fig. 3.